\documentclass[aps,prd,onecolumn,eqsecnum,amsmath,nofootinbib,preprintnumbers]{revtex4}

\usepackage[dvipsnames]{xcolor}
\usepackage{color,graphicx,float,subfigure,xcolor}
\usepackage{amsfonts,amssymb,theorem,mathrsfs,times}
\usepackage{bm}
\usepackage{multirow}
\usepackage{mathtools}
\usepackage{amsfonts,amssymb,theorem,mathrsfs}
\usepackage{dsfont}
\usepackage{setspace}
\usepackage{amsmath} 
\usepackage{graphicx}
\usepackage[colorlinks,linkcolor=blue,anchorcolor=blue,citecolor=blue]{hyperref}
\usepackage{ulem}
\usepackage{booktabs} 

\begin{document}
\title{Total transmission modes in draining bathtub model with vorticity}
\date{\today}

\author{Zhe Yu$^{a}$\footnote{e-mail address: yuzhe@nbu.edu.cn}} 

\author{Liang-Bi Wu$^{b\, ,c}$\footnote{e-mail address: wulb@ucas.ac.cn (corresponding author)} }

\affiliation{${}^a$Institute of Fundamental Physics and Quantum Technology, Ningbo University, Ningbo 315211, China}

\affiliation{${}^b$Interdisciplinary Center for Theoretical Study and Department of Modern Physics,\\
University of Science and Technology of China, Hefei, Anhui 230026, China}

\affiliation{${}^c$School of Fundamental Physics and Mathematical Sciences, Hangzhou Institute for Advanced Study, UCAS, Hangzhou 310024, China}

\begin{abstract}
We investigate the total transmission modes (TTMs) in the draining bathtub model (DBM) with vorticity using the Chebyshev-Lobatto pseudospectral method, where the boundary conditions of the total transmission modes are both ingoing at the event horizon and infinity. Numerical results show that the (right) TTM spectra can possess positive imaginary parts, while for certain parameters they acquire negative imaginary parts. The extreme sensitivity of the higher overtones is manifested as their pronounced spectral mobility.
\end{abstract}

\maketitle

\section{Introduction}
Quasinormal modes (QNMs) are a discrete set of damped resonances that characterize the relaxation phase of perturbed open systems~\cite{Nollert:1999ji,Kokkotas:1999bd,Berti:2025hly,Berti:2009kk,Trivedi:2026qof}. In gravitational physics, these modes govern the late-time ringdown of a perturbed black hole, such as those formed following binary mergers. According to the no-hair theorem, the complex frequencies of QNMs are determined exclusively by the intrinsic parameters of the black hole, such as its mass and angular momentum. Consequently, QNMs serve as unique characteristic fingerprints of the background spacetime, making their spectrum a powerful tool for probing the strong-field regime of gravity, extracting black hole parameters, and testing general relativity~\cite{LIGOScientific:2016lio,LIGOScientific:2020tif,LIGOScientific:2021sio}.

From the perspective of scattering theory, in addition to the QNMs which typically correspond to poles of scattering amplitudes, there exists another class known as the total transmission modes (TTMs). In contrast to standard QNMs that govern the late-time relaxation of an open system, TTMs constitute a unique class of complex-frequency resonances that can be dynamically excited through suitably tailored time-dependent scattering processes~\cite{Chandrasekhar:1985kt,Andersson:1994tt,MaassenvandenBrink:2000iwh,Berti:2004md,Keshet:2007be,Cook:2016fge,Cook:2016ngj,Cook:2018ses,Cook:2022kbb}. When an incoming wave packet is specifically modulated to match the complex frequencies of these modes, the system exhibits the remarkable phenomenon of ``virtual absorption''~\cite{Longhi:2010amr}. During this process, the reflected signal is entirely suppressed for the full duration of the wave-potential interaction, effectively rendering the effective potential perfectly transparent and making the compact object act as a perfect absorber~\cite{Tuncer:2025dnp,Wu:2026hvf}. This property makes TTMs an exceptionally rigorous probe for the underlying spacetime geometry and possibly of quantum properties of the black hole~\cite{Keshet:2007be,Kwon:2011zza}.  Furthermore, recent investigations into the spectral stability of these exotic resonances have revealed that TTMs, much like their QNM counterparts, are generically sensitive to small environmental perturbations~\cite{Zhou:2025xdo}. Analysis utilizing pseudospectrum techniques~\cite{Jaramillo:2020tuu,Cao:2024sot,Cao:2024oud,Destounis:2021lum,Chen:2024mon,Cai:2025irl,Cao:2025qws} and condition numbers demonstrates that the spectral instability of TTMs increases for higher overtones, underscoring their highly non-trivial dependence on the details of the effective potential. Consequently, exploring the existence, stability, and parameter dependence of TTMs across diverse effective spacetimes has emerged as a compelling frontier in modern wave dynamics.

Despite their profound theoretical implications, the direct observation of TTMs in astrophysical black holes remains particularly challenging. The specific temporal modulations required for their excitation, together with their extreme sensitivity to environmental perturbations, make these modes exceedingly difficult to detect in realistic, noisy astrophysical settings. As an alternative to direct astrophysical observation, analogue gravity offers a powerful platform for probing such exotic resonant phenomena in a controlled laboratory environment~\cite{Barcelo:2005fc}. Actually, analogue gravity is an approach that exploits the fundamental mathematical equivalence between the propagation of small perturbations in certain physical systems and the dynamics of fields in curved spacetimes~\cite{Barcelo:2005fc}. Over the past decades, this cross-disciplinary paradigm has been extensively investigated through a rich variety of physical setups, prominently including acoustic perturbations in classical fluids~\cite{Unruh:1980cg,Visser:1997ux}, surface waves in shallow water~\cite{Schutzhold:2002rf,Weinfurtner:2010nu}, Bogoliubov excitations in Bose-Einstein condensates~\cite{Garay:1999sk,Barcelo:2000tg}, and electromagnetic waves in complex optical media~\cite{Philbin:2007ji,Belgiorno:2010wn}. These analogue platforms enable the investigation of non-perturbative scattering processes and fundamental wave phenomena that are otherwise inaccessible in realistic astrophysical environments. For instance, they have been successfully employed to simulate Hawking emission near acoustic horizons~\cite{Steinhauer:2015saa} and to experimentally demonstrate Penrose superradiance in fluid vortex flows~\cite{Torres:2016iee}. Among these, hydrodynamic systems are of particular importance, with the draining bathtub model (DBM) serving as a canonical acoustic analogue for rotating black holes~\cite{Visser:1997ux}. In the absence of vorticity, this effective stationary spacetime has enabled extensive studies of the fundamental QNMs and characteristic ringing of analogue black holes~\cite{Berti:2004ju,Dolan:2009zza}. Crucially, introducing vorticity endows the propagating scalar field with a local effective mass, fundamentally altering the potential configuration and leading to long-lived quasibound states~\cite{Patrick:2018orp,Patrick:2019prh}. More recently, spectrum instability of DBM is studied in~\cite{MalatoCorrea:2025iuc,dePaula:2025fqt}.

Motivated by the influence of background vorticity on local wave dynamics, this work systematically investigates the TTMs within the DBM. We demonstrate that the vorticity-induced modification of the effective potential barrier can naturally give rise to these exact total transmission resonances. Given the strong sensitivity of TTM to the detailed spatial profile of the effective potential, the DBM offers an ideal and highly tunable platform for their study. Investigating TTMs in this hydrodynamic vortex system not only connects the theoretical concept of virtual absorption~\cite{Tuncer:2025dnp,Wu:2026hvf} to experimental realizability but also provides a rigorous test of the robustness of these anomalous resonance phenomena in complex rotating effective geometries.

The remainder of this paper is organized as follows. In Sec. \ref{sec:set}, we introduce the methodology for solving the wave equation in the frequency domain for the DBM with vorticity. In Sec. \ref{sec: TTM}, we analyze the numerical results and discuss the corresponding (right) TTM spectral characteristics. Finally, Sec. \ref{sec:conclusion} is devoted to our conclusions. Additional technical details are provided in the appendices. Specifically, Appendix~\ref{app:WaveEquation} details the derivation of the wave equation with vorticities under the shallow water approximation, while Appendix~\ref{sec:coefficients} presents the explicit coefficients required for the Chebyshev-Lobatto grid.

\section{Set up}\label{sec:set}
In this section, we establish a theoretical framework to systematically calculate the TTMs in the DBM. A brief review of the DBM and the derivation of the wave equation incorporating vorticity are provided in Appendix \ref{app:WaveEquation}. Specifically, the wave equation for the DBM in the frequency domain  can be expressed as~\cite{Patrick:2018orp}
\begin{eqnarray}\label{master_equation}
    \Big[\frac{\mathrm{d}^2}{\mathrm{d}x^2}-V_{\text{eff}}(r(x))\Big]\Psi(x)=0\, ,
\end{eqnarray}
where $x$ denotes the tortoise coordinate. The relationship between $x$ and $r$ is specified by
\begin{eqnarray}\label{relation_x_and_r}
    \mathrm{d}x=\frac{c\mathrm{d}r}{c^2-V_r^2}\, ,\quad V_r=-\frac{D}{r}\, ,
\end{eqnarray}
where $c>0$ is the speed of sound and $V_r$ is the radial velocity. By integrating Eqs. \eqref{relation_x_and_r}, the coordinate $x$ can be explicitly derived as 
\begin{eqnarray}\label{tortoise_coordinate}
    x=\frac{r}{c}+\frac{D}{2c^2}\ln\Big[\frac{2c(cr-D)}{D+cr}\Big].
\end{eqnarray}
In Eq. \eqref{master_equation}, the effective potential $V_{\text{eff}}$ in terms of the radial coordinate $r$ is given by 
\begin{eqnarray}\label{eff_potential}
    V_{\mathrm{eff}}(r)=-\Big(\omega-\frac{mV_{\theta}}{r}\Big)^2+\Big(c^2-\frac{D^2}{r^2}\Big)\Big(\frac{m^2-1/4}{r^2}+\frac{5D^2}{4c^2r^4}+\frac{\Omega_v^2}{c^2}\Big) \, ,
\end{eqnarray}
where $\Omega_v$ represents the local vorticity of the fluid and can be written as
\begin{eqnarray}\label{vorticity}
    \Omega_v=\frac{1}{r}\frac{\partial(rV_\theta(r))}{\partial r}.
\end{eqnarray}
Following Ref.~\cite{Patrick:2018orp}, we consider a natural vortex profile of the form
\begin{eqnarray}
    V_\theta(r)=\frac{Br}{r_0^2+r^2}\, .
\end{eqnarray}
This profile of vortex is characterized by two distinct physical regimes. In the core region where $r$ is satisfied with $r \ll r_0$, the velocity for the angular part scales as $V_\theta \propto r$, representing a rotational flow dominated by vorticity. Conversely, in the far-field limit namely $r \gg r_0$, the velocity follows $V_\theta \propto 1/r$, which recovers the irrotational behavior typical of a free vortex.

Now, we are going to solve Eq. \eqref{master_equation} by imposing boundary conditions at event horizon and infinity. In contrast to the boundary of the QNMs, there exist only ingoing waves at both boundaries, so the conditions of (right) TTMs have the following form~\cite{Tuncer:2025dnp,Zhou:2025xdo}, 
\begin{align}\label{boundary_condition}
    \Psi(x) \sim 
    \begin{cases} 
    \mathrm{e}^{-\mathrm{i}(\omega - m\Omega_\text{H})x} & x \to -\infty \quad  \\[10pt]
    \mathrm{e}^{-\mathrm{i} \omega x} & x \to \infty \quad 
    \end{cases}\, .
\end{align}
By adopting the above effective potential profile, it can be proved that the horizon $r_{\text{H}}=D/c$ is a regular singular point of Eq. \eqref{master_equation}. Following the standard method to obtain series solution near a regular singular point~\cite{wang1989special}, one can get the two indices of the indicial equation
\begin{eqnarray}\label{indices_at_horizon}
    \rho_{1,2}=\pm\mathrm{i}\frac{D}{2c^2}\Big(\omega-m\Omega_\text{H}\Big)\, ,
\end{eqnarray}
where $\Omega_\text{H}$ is the angular frequency at the horizon and its specific form is
\begin{eqnarray}\label{angular_frequency_at_horizon}
    \Omega_\text{H}=\Big(\frac{V_{\theta}}{r}\Big)\Bigg|_{r=r_\text{H}}=\frac{B c^2}{D^2+r_0^2c^2}\, .
\end{eqnarray}
Combining with the above boundary condition \eqref{boundary_condition}, one finds that only the index $\rho_2$ is admissible for right TTMs. In addition, $r=\infty$ can be proved to be an irregular singular point. The asymptotic behavior of the solution in the vicinity of this singularity can be characterized by the standard form $\exp[Q(r)]w(r)$. A detailed analysis reveals that $Q(r) = -\mathrm{i}\omega r/c$, while $w(r)$ represents the regular part of the solution and admits a Frobenius-like expansion $w(r) = \sum_{n=0}^\infty a_n r^{-n}$. As a result, the solution of Eq. \eqref{master_equation} in terms of the radial coordinate $r$ takes the following ansatz
\begin{eqnarray}\label{ansatz}
    \Psi(r)=\Big(1-\frac{r_{\text{H}}}{r}\Big)^{-\frac{\mathrm{i}D}{2c^2}(\omega-m\Omega_\text{H})}\mathrm{e}^{-\mathrm{i}\frac{\omega}{c}r}\tilde{\Psi}(r)\, ,
\end{eqnarray}
where $\tilde{\Psi}$ is a regular function. The above ansatz ensures that solutions at the horizon and infinity are both ingoing. For applying pseudospectral method~\cite{Jaramillo:2020tuu,Jansen:2017oag,Cai:2025irl}, Eq. \eqref{master_equation} should be rewritten in terms of the compact coordinate $\sigma$ where $\sigma=1-r_{\text{H}}/r$. Substituting Eq. \eqref{ansatz} into Eq. \eqref{master_equation}, one finally gets
\begin{eqnarray}\label{numerical_Eq}
    \Bigg\{C_2(\sigma)\frac{\mathrm{d}^2}{\mathrm{d}\sigma^2}+\Big[C_1^{[0]}(\sigma)+C_1^{[1]}(\sigma)\omega\Big]\frac{\mathrm{d}}{\mathrm{d}\sigma}+\Big[C_0^{[0]}(\sigma)+C_0^{[1]}(\sigma)\omega+C_0^{[2]}(\sigma)\omega^2\Big]\Bigg\}\tilde{\Psi}(\sigma)=0\, ,
\end{eqnarray}
where the expressions of the coefficients $C_2(\sigma)$, $C_1^{[0]}(\sigma)$, $C_1^{[1]}(\sigma)$, $C_0^{[0]}(\sigma)$, $C_0^{[1]}(\sigma)$ and $C_0^{[2]}(\sigma)$ are displayed in Appendix~\ref{sec:coefficients}, and the conditions $C_2(1)=C_2(0)=0$ are satisfied at the boundaries. Using the Chebyshev-Lobatto (CL) grid, the above equation (\ref{numerical_Eq}) can be discretized into
\begin{eqnarray}\label{T_omega_problem}
    \mathbf{T}(\omega)\tilde{\mathbf{\Psi}}=0\, ,
\end{eqnarray}
where the matrix $\mathbf{T}(\omega)$ reads
\begin{eqnarray}
\label{general_eigenvalue_problem}
    \mathbf{T}(\omega)&=&\Big(\mathbf{C}_2\cdot\mathbf{D}^2+\mathbf{C}_1^{[0]}\cdot\mathbf{D}+\mathbf{C}_0^{[0]}\Big)+\omega\Big(\mathbf{C}_1^{[1]}\cdot\mathbf{D}+\mathbf{C}_0^{[1]}\Big)+\omega^2\mathbf{C}_0^{[2]}\nonumber\\
    &\equiv&\mathbf{T}_0+\mathbf{T}_1\omega+\mathbf{T}_2\omega^2\, .    
\end{eqnarray}
and $\mathbf{D}$ refers to the matrix representation of the differential operator on such grid. Following the approach used in~\cite{Jansen:2017oag,Li:2024npg,Li:2025ljb,Xie:2025jbr}, the polynomial eigenvalue problem for the matrix $\mathbf{T}(\omega)$ can be transformed into a generalized eigenvalue problem. By introducing the auxiliary variable $\tilde{\mathbf{\Phi}}= \omega\tilde{\mathbf{\Psi}}$, Eq.~\eqref{T_omega_problem} can be recast into the following block matrix form
\begin{eqnarray}\label{general_eigenvalue_problem}
    \begin{pmatrix} 
    \mathbf{T}_0 & \mathbf{T}_1\\
    \mathbf{0} & \mathbf{I}   
    \end{pmatrix}
\begin{pmatrix} 
\tilde{\mathbf{\Psi}} \\ 
\tilde{\mathbf{\Phi}} 
\end{pmatrix}
= \omega
    \begin{pmatrix} 
    \mathbf{0} & -\mathbf{T}_2\\
    \mathbf{I} & \mathbf{0}   
    \end{pmatrix} 
    \begin{pmatrix} 
        \tilde{\mathbf{\Psi}} \\ 
        \tilde{\mathbf{\Phi}} 
    \end{pmatrix}\, .
\end{eqnarray}
The TTM spectra are then obtained by solving the derived generalized eigenvalue problem, which can be efficiently evaluated using the built-in function \texttt{Eigenvalues} in \textit{Mathematica}. To ensure spectral convergence, the eigenvalues are calculated by using different resolutions. The results are deemed converged when the relative difference between eigenvalues obtained at increasing matrix dimensions is less than $10^{-7}$.

\section{TTM spectra}\label{sec: TTM}
In this section, we present the numerical results for TTM spectra in the DBM, obtained by solving the generalized eigenvalue problem derived in Eq. \eqref{general_eigenvalue_problem}. For numerical convenience and without loss of generality, we adopt dimensionless units by setting $c=1$ and $D=1$. This fixes the acoustic horizon at $r_{\text{H}}=1$, allowing us to naturally isolate the intrinsic resonance properties and their dependence on the vorticity parameters $B$ and $r_0$. 
\begin{table}[htbp]
\centering
\renewcommand{\arraystretch}{1.5} 
\begin{tabular}{ccccc}
\hline\hline
$m$~~~& $n=0$ ~~~& $n=1$ ~~~& $n=2$ ~~~& $n=3$ \\ \hline\noalign{\vskip 5pt}
1 ~~~& $-0.3301 + 0.0517\mathrm{i}$~~~& $-9.2416 + 8.0526\mathrm{i}$~~~& $14.0285 + 9.0596\mathrm{i}$~~~& $-11.0341 + 20.5954\mathrm{i}$\\ \hline\noalign{\vskip 5pt}
2 ~~~& $-0.5229 - 0.4023\mathrm{i}$ ~~~& $-0.4158 + 1.0587\mathrm{i}$ ~~~& $- 6.7388+9.2362\mathrm{i}$   ~~~& $16.5471 + 11.4722\mathrm{i}$  \\ \hline\noalign{\vskip 5pt}
3 ~~~& $-0.5907 - 0.8684\mathrm{i}$~~~& $-1.4575 + 0.5457\mathrm{i}$~~~& $ 18.9392 + 14.0596\mathrm{i}$~~~& $-5.2176 + 32.8025\mathrm{i}$  \\ \hline\hline\noalign{\vskip 5pt}
\end{tabular}
\caption{The first four TTM spectra for $m=1, 2, 3$, where $n$ denotes the overtone number. The parameters are adopted as $B=2.5$, $r_0=1.2$.}
\label{tab:ttm}
\end{table}

\begin{table}[htbp]
\centering
\renewcommand{\arraystretch}{1.5}
\begin{tabular}{ccccc}
\hline\hline
$m$~~~& $n=0$~~~& $n=1$~~~& $n=2$~~~& $n=3$ \\ \hline\noalign{\vskip 5pt}
1~~~& $-0.2416 + 0.2490\mathrm{i}$~~~& $-3.8457 + 5.0194\mathrm{i}$~~~& $6.4045 + 5.6738\mathrm{i}$~~~& $-4.8496 + 12.5195\mathrm{i}$\\ \hline\noalign{\vskip 5pt}
2~~~& $-0.5182 + 0.1506\mathrm{i}$~~~& $-2.4083 + 6.0850\mathrm{i}$~~~& $7.7959 + 7.3354\mathrm{i}$~~~& $9.5880 + 14.5904\mathrm{i}$  \\ \hline\noalign{\vskip 5pt}
3~~~& $-0.7263 + 0.0553\mathrm{i}$~~~& $-0.5467 + 1.0745\mathrm{i}$~~~& $9.0957 + 8.9995\mathrm{i}$~~~& $11.0664 + 16.1106\mathrm{i}$ \\ \hline\hline\noalign{\vskip 5pt}
\end{tabular}
\caption{Another set of TTM spectra for $m=1, 2, 3$. The parameters are chosen as $B=2$, $r_0=1.5$.}
\label{tab:ttm2}
\end{table}

Since we aim to elucidate the evolution of the TTM spectra with varying parameter $r_0$ or $B$, it is necessary to specify the fiducial reference position for the spectral migration. Tables \ref{tab:ttm} and \ref{tab:ttm2} summarize some results of the TTM spectra for the fundamental mode ($n=0$) and the first three overtones ($n=1, 2, 3$) with azimuthal numbers $m=1, 2, 3$ under different vortex profiles. Specifically, Table \ref{tab:ttm} assumes $B=2.5$ and $r_0=1.2$, whereas Table \ref{tab:ttm2} considers $B=2$ and $r_0=1.5$. The imaginary parts of some modes in Table \ref{tab:ttm} are negative while all modes listed in \ref{tab:ttm2} have positive imaginary parts. The overtones are ordered according to the signed magnitude of their imaginary parts. A direct comparison of the tabulated data reveals that the TTM spectra are highly sensitive to both the azimuthal number $m$ and the overtone number $n$. 

\begin{figure}
\includegraphics[width=0.48\textwidth]{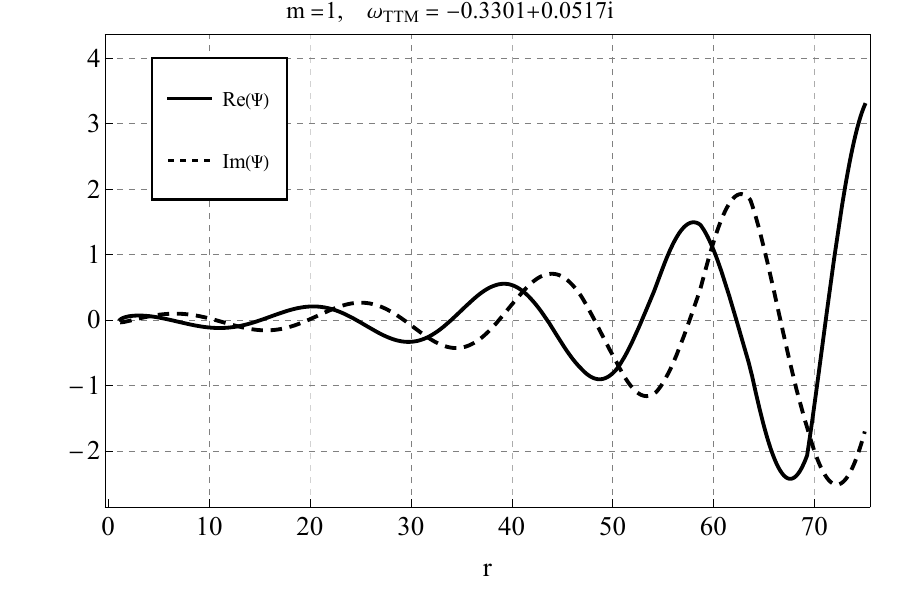} \hfill
\includegraphics[width=0.48\textwidth]{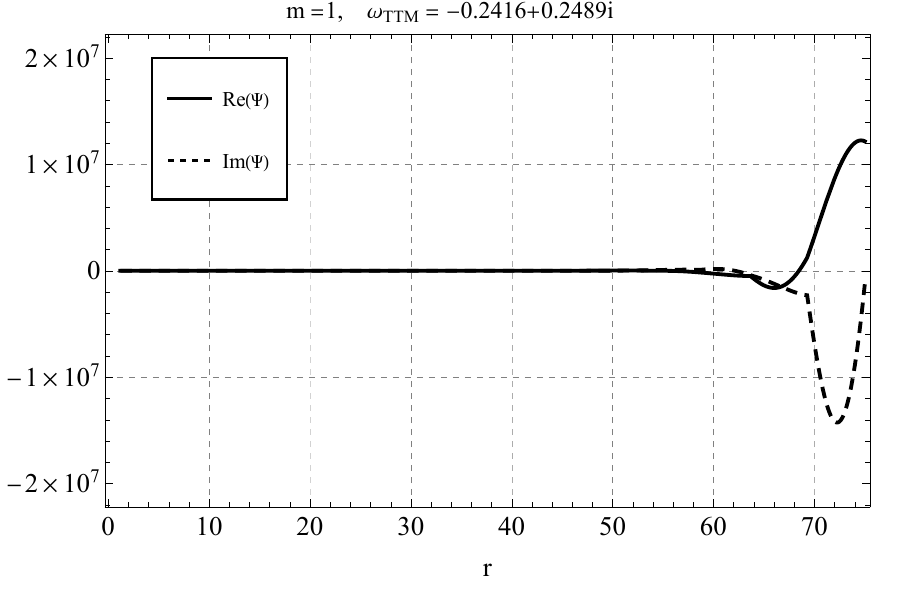} \\
\vspace{1em} 
\includegraphics[width=0.48\textwidth]{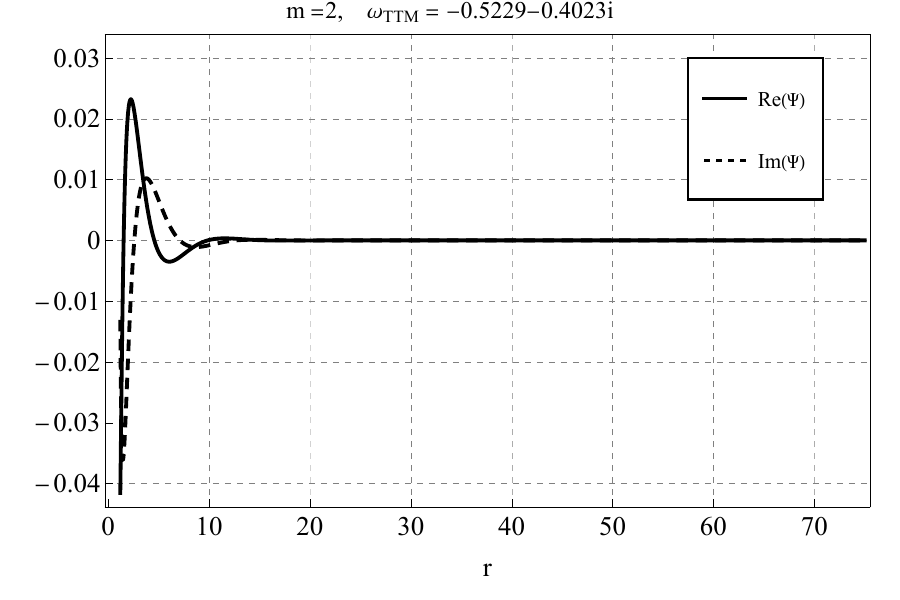} \hfill
\includegraphics[width=0.48\textwidth]{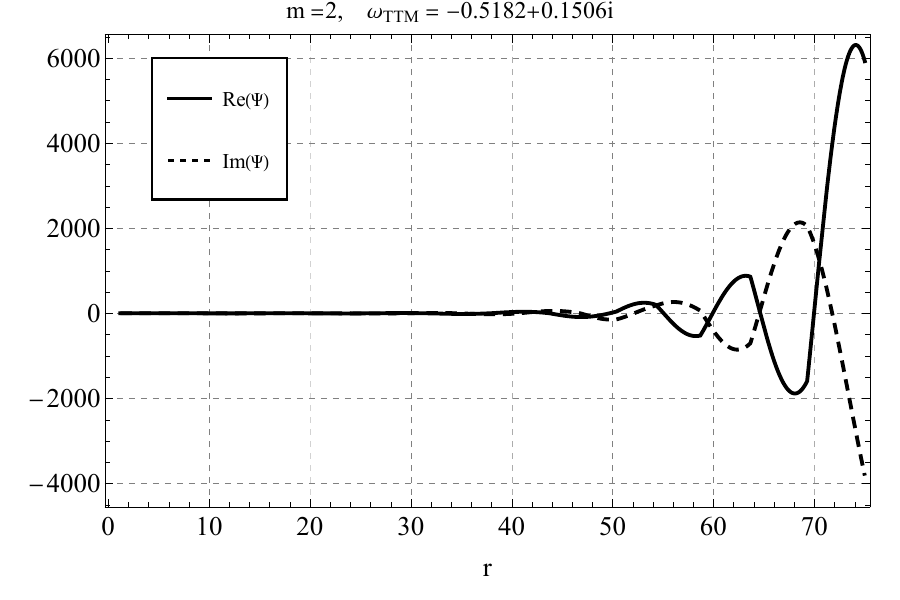} \\
\vspace{1em} \includegraphics[width=0.48\textwidth]{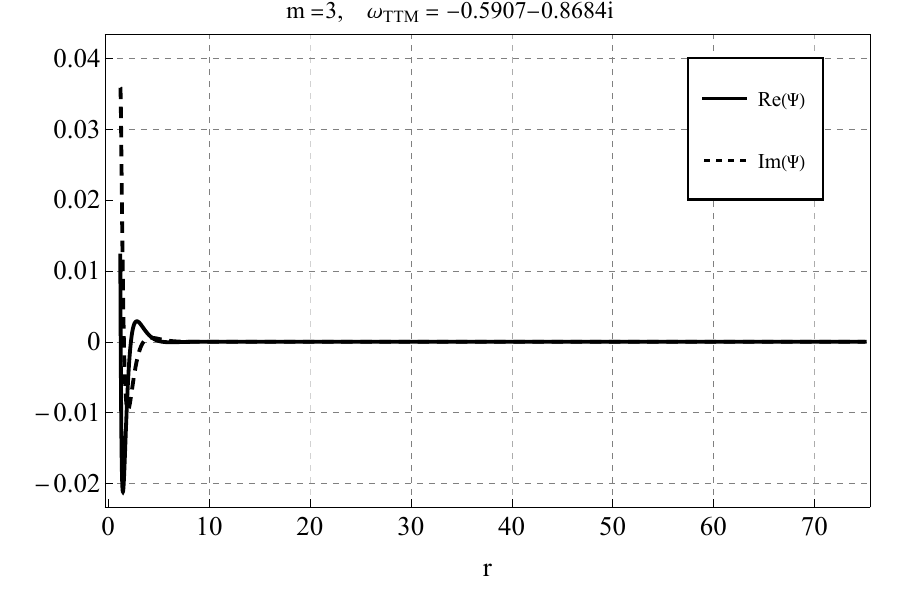} \hfill
\includegraphics[width=0.48\textwidth]{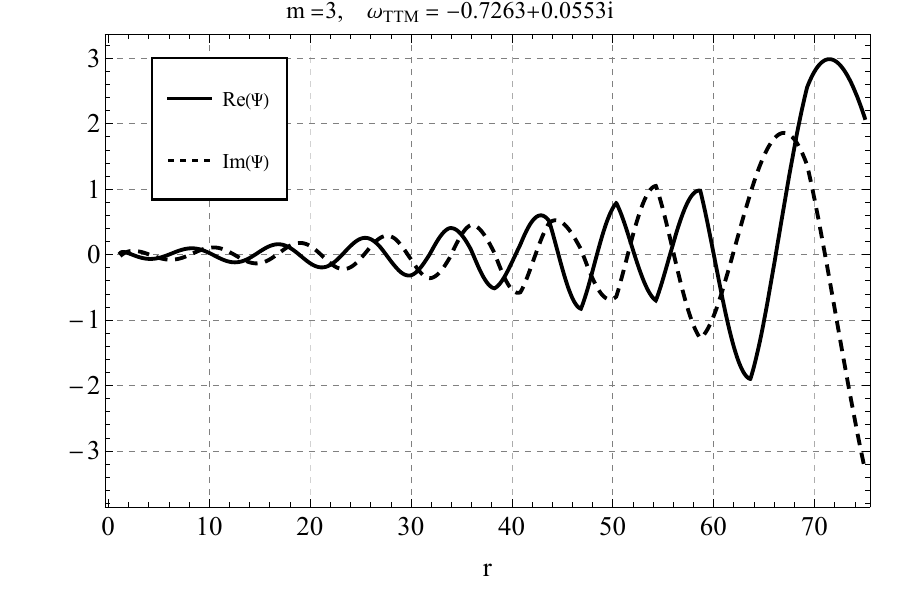}
\caption{Eigenfunctions of the fundamental modes for $m=1, 2, 3$. The modes shown in the left and right panels correspond to the data selected in Tables \ref{tab:ttm} and \ref{tab:ttm2}, respectively.  In all panels, the radial coordinate ranges from $r\in[1.2, 74]$.}
\label{fig1}
\end{figure}

For the completeness of the results, we not only present the results of spectra, but also the corresponding eigenfunctions. Fig.~\ref{fig1} displays the radial profiles of the eigenfunctions for the modes selected in the Tables \ref{tab:ttm} and \ref{tab:ttm2} with $m=1, 2, 3$, in which we first obtain the eigenfunctions $\tilde{\Psi}$ in CL grid and get the true eigenfunction $\Psi$ in terms of the coordinate $r$ via Eq. \eqref{ansatz}. Each panels in Fig. \ref{fig1} illustrate the real and imaginary parts of the eigenfunctions $\Psi$, where solid lines stand for real parts and dashed lines stand for imaginary parts. These results demonstrating distinct oscillatory behaviors near the vortex core and their adherence to the specified incoming boundary conditions \eqref{boundary_condition}. Specifically, if the imaginary part of $\omega$ is positive, the wave function diverges as $r\to \infty$, whereas if it is negative, the wave function diverges as $r\to r_{\text{H}}$. The deformations in the wave profiles across different azimuthal modes are primarily driven by the rotational Doppler shift, which corresponds to the first term of the effective potential \eqref{eff_potential}. This effect locally shifts the spectra, thereby modulating the potential landscape. Consequently, the radial distribution of the wave function must adjust to the modified potential barriers, physically manifesting as the observed variations in the wave shapes. Moreover, TTMs are absent for $m=0$. In this case, the effective potential \eqref{eff_potential} reduces to the spherically symmetric case, where the lack of a potential barrier inhibits the excitation of TTMs.

\begin{figure}
\includegraphics[width=0.48\textwidth]{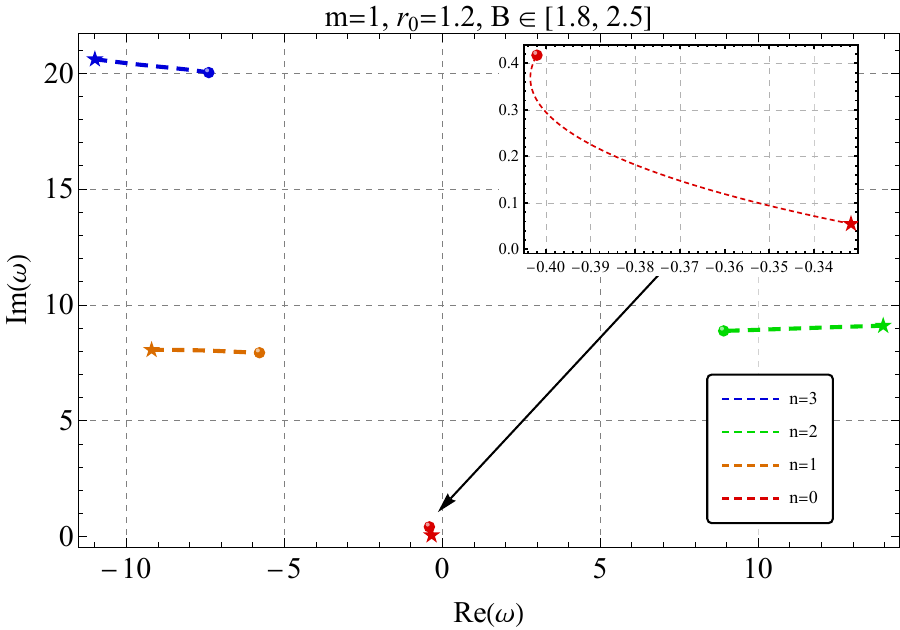} \hfill
\includegraphics[width=0.48\textwidth]{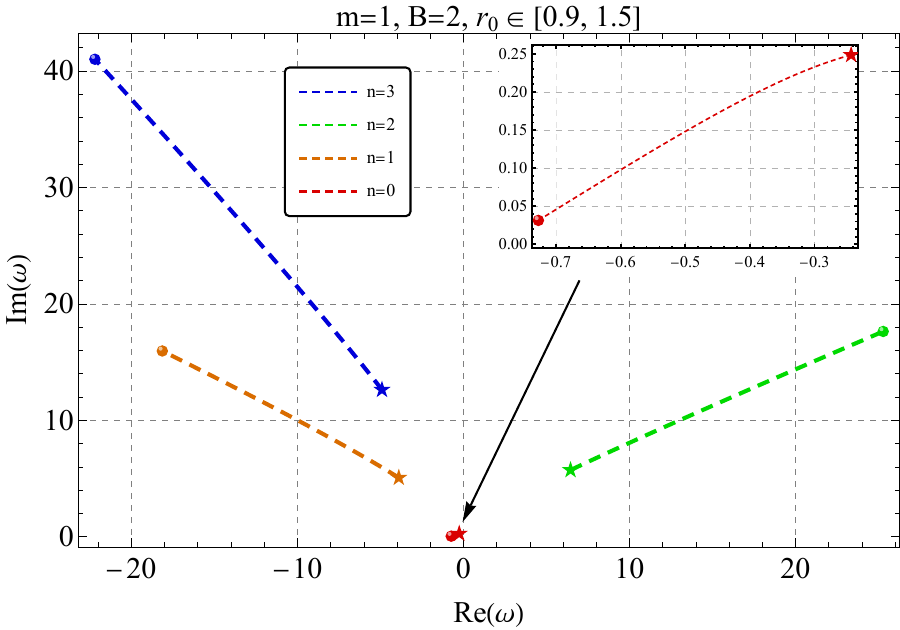} \\
\vspace{1em} 
\includegraphics[width=0.48\textwidth]{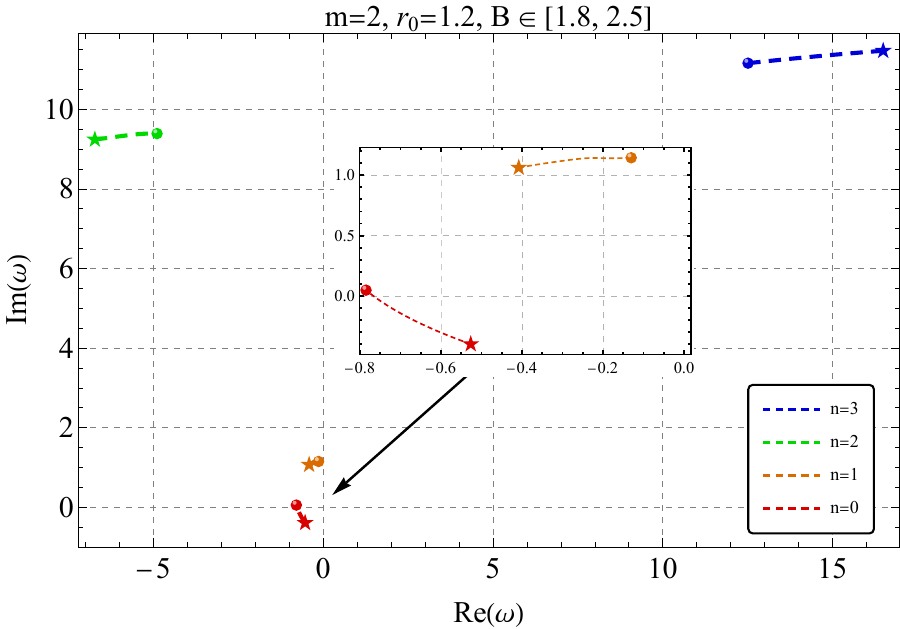} \hfill
\includegraphics[width=0.48\textwidth]{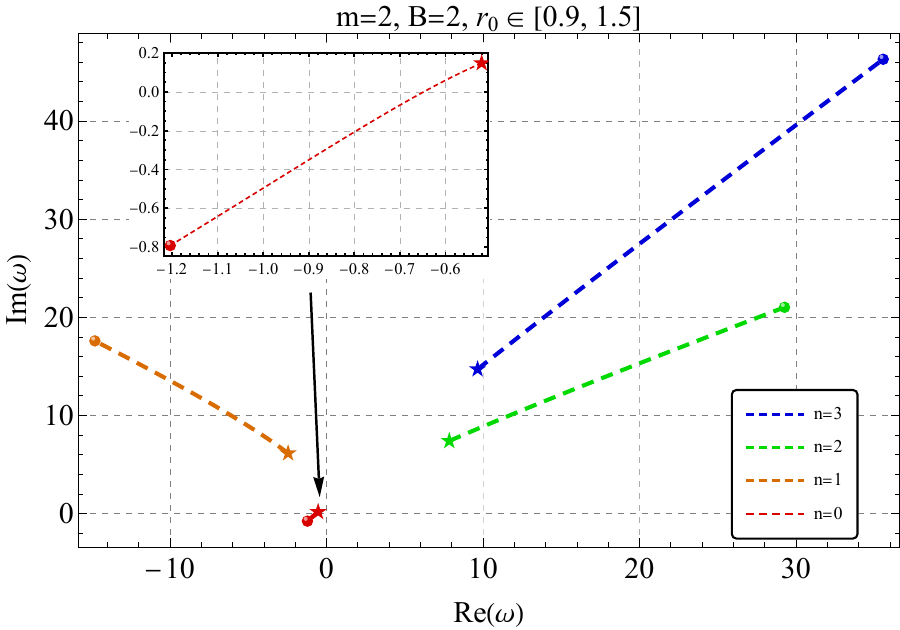} \\
\vspace{1em} \includegraphics[width=0.48\textwidth]{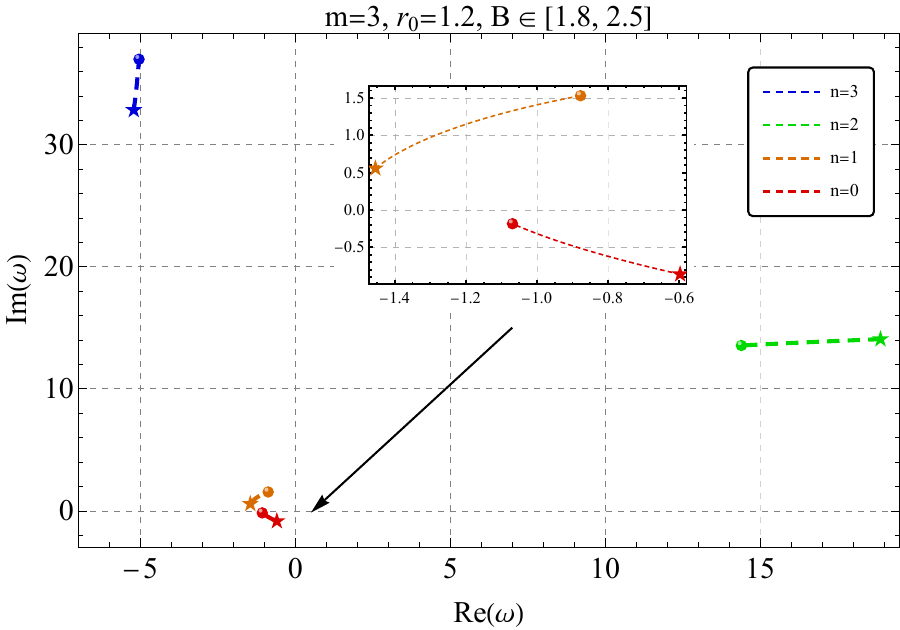} \hfill
\includegraphics[width=0.48\textwidth]{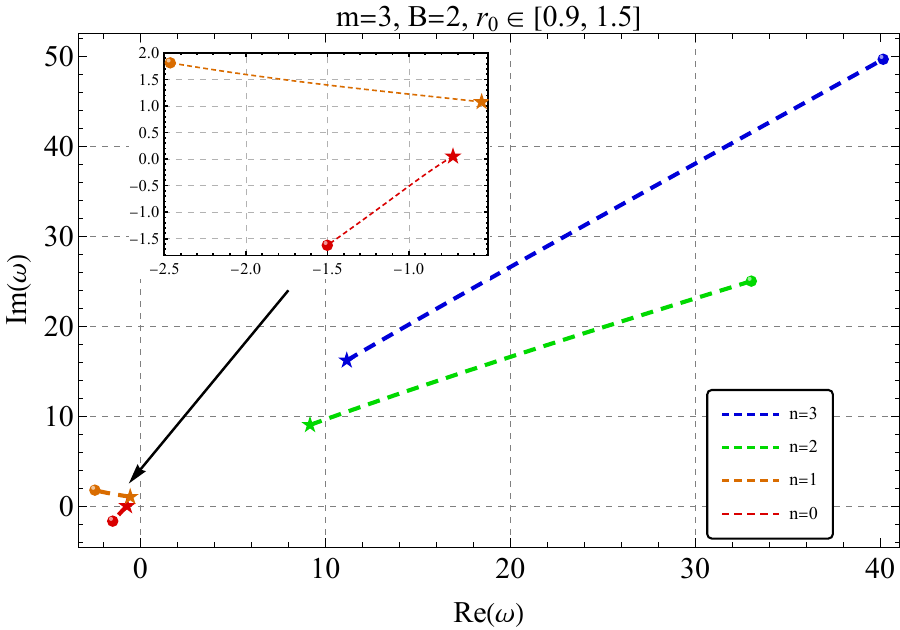}
\caption{The left panels display the TTM trajectories as the parameter $B$ varies in the range $[1.8, 2.5]$ for $m=1, 2, 3$ from top to bottom, with a fixed $r_0=1.2$. The right panels illustrate the corresponding trajectories as the parameter $r_0$ varies in the range $[0.9, 1.5]$ for $m=1, 2, 3$ from top to bottom, with a fixed $B=2$. Each panel presents four distinct trajectories corresponding to $n=0, 1, 2, 3$. For every individual trajectory, a solid circle and a star indicate the starting and ending points of the parameter variation, respectively.  Note that for a fluid with reversed vorticity where $B<0$, the TTM trajectories exhibit reflection symmetry across the imaginary axis relative to the results shown here.}
\label{fig2}
\end{figure}

To further investigate migrations of these TTM spectra with respect to some physical parameters, we trace their trajectories in the complex $\omega$ plane under continuous variations of the background flow parameters, as illustrated in Fig. \ref{fig2}. The trajectory patterns highlight how the deformation of the effective potential $V_{\text{eff}}(r)$, driven by shifts in the rotation strength $B$ and the core radius $r_0$, inevitably causes the migration of the TTMs. Notably, we observe a stark contrast in the migration behavior between the fundamental mode and the higher overtones. The trajectory lengths for higher overtones are remarkably more extended, visually demonstrating that they are considerably more susceptible to small parametric shifts in the background. In addition, we can also see from Fig. \ref{fig2} that there are some parameter ranges that cause the TTM spectrum to migrate from the upper half plane to the lower half plane. Furthermore, it is mathematically straightforward to verify that for a fluid with reversed vorticity where $B<0$, the corresponding TTM spectra trajectories exhibit strict reflection symmetry across the imaginary axis.

Ultimately, these numerical results confirm that the vorticity-induced modification of the effective potential in the DBM naturally gives rise to these exact total transmission resonances. The pronounced spectral mobility of the higher overtones perfectly aligns with recent theoretical findings regarding the spectrum instability of TTMs~\cite{Zhou:2025xdo}, reinforcing their potential as exceptionally rigorous probes for the underlying geometry of analogue black holes.

\section{Conclusions}\label{sec:conclusion}
Because direct observations of gravitational scattering around astrophysical black holes remain formidably challenging, analog gravity paradigm provides a crucial, laboratory-accessible alternative for probing these complex dynamics. Motivated by this framework, in the frequency domain, we investigate the TTMs using the DBM as a canonical fluid-mechanical analogue.

By rigorously formulating the scalar wave equation with background vorticity, we mapped the physical problem into a generalized eigenvalue framework. Employing the highly accurate Chebyshev-Lobatto pseudospectral method, we successfully extracted the TTM spectra and corresponding eigenfunctions for both the fundamental modes and higher overtones. Furthermore, our trajectory analysis in the complex plane visually demonstrated the pronounced spectral mobility of these exotic resonances. Tracking their migration confirmed that higher overtones are remarkably more sensitive to continuous variations in the background vortex parameters than the fundamental modes. This heightened sensitivity underscores their highly non-trivial dependence on the details of the effective potential, highlighting the rich scattering structure inherent in rotating spacetimes.

Despite the robust theoretical presence of TTMs in the frequency domain, achieving exact dynamic suppression of reflected signals in the time domain remains nontrivial due to the existence of superradiant amplification of rotating backgrounds. A detailed analysis of TTMs in the time-domain framework will be addressed in future work.

\appendix

\section*{Acknowledgement}
We thanks for Yu-Sen Zhou for helpful discussion. This work is supported by the National Natural Science Foundation of China. Specifically, Zhe Yu is supported by Grant No. 12447176, and Liang-Bi Wu is supported by Grant No. 12505067.

\appendix
\section{The wave equation with vorticity in DBM}
\label{app:WaveEquation}
In this appendix, we briefly review the draining bathtub model (DBM). We then derive the wave equation that incorporates vorticity. Our derivation follows the approach in Ref. \cite{Patrick:2019prh} and can also be seen in Ref. \cite{Barcelo:2005fc}.

Let $\boldsymbol{\nabla}^{(3)}$ denote the three-dimensional gradient operator, and let $\boldsymbol{\nabla}$ represent the gradient operator in the two-dimensional plane. Similarly, the three-dimensional velocity field is denoted by $\mathbf{V}^{(3)}=\big(\tilde{\mathbf{V}},V_z\big)$, where $\tilde{\mathbf{V}}$ represents the horizontal velocity components and $V_z$ is the vertical component. An irrotational, incompressible, and inviscid fluid is useful for mimicking gravity. The Euler equation and continuity equation for such kind of fluid are
\begin{eqnarray}\label{Euler_Eq}
    \partial_t \mathbf{V}^{(3)}+\tilde{\mathbf{V}}\cdot\boldsymbol{\nabla}\mathbf{V}^{(3)}+V_z\partial_z\mathbf{V}^{(3)}+\frac{1}{\rho} \boldsymbol{\nabla}^{(3)} P=\mathbf{g}\, ,
\end{eqnarray}
and
\begin{eqnarray}\label{continuity_Eq}
    \boldsymbol{\nabla}\cdot\tilde{\mathbf{V}}+\partial_z V_z=0\, ,
\end{eqnarray}
where $\mathbf{g}=\left(0,0,-g\right)$ is the gravitational acceleration of the Earth, and other forces are neglected.

In the DBM, the shallow water approximation is crucial. This approximation implies that the fluid height $\tilde{H}$ is much smaller than the horizontal scale $L$, i.e., $\tilde{H}\ll L$ and $V_z\ll |\tilde{\mathbf{V}}|$. Consequently, the fluid can be effectively treated as two-dimensional, leading to a $(2+1)$-dimensional analogue spacetime metric.

By virtue of this approximation, the $z$-component of Eq. \eqref{Euler_Eq} can be written as
\begin{eqnarray}
    \frac{1}{\rho} \partial_z P=-g\, ,
\end{eqnarray}
where vertical hydrostatic equilibrium is assumed, since $V_z$ is small enough to be treated as a perturbation. By restricting $\tilde{\mathbf{V}}$ to be independent of $z$ and integrating Eq.~\eqref{continuity_Eq} over the vertical coordinate, we obtain
\begin{eqnarray}\label{V_and_P}
    \begin{aligned}
    V_z=-z \boldsymbol{\nabla} \cdot \tilde{\mathbf{V}}\, ,\qquad
    P=\rho g(\tilde{H}-z)\, ,
    \end{aligned}
\end{eqnarray}
where $\tilde{H}$ is the fluid height, and the no-penetration boundary condition $V_z(z=0)=0$ along with the free-surface boundary condition $P(z=\tilde{H})=0$ have been imposed. By adopting the definition $V_z=\mathrm{d}z/\mathrm{d}t$ which yields $V_z(z=\tilde{H})=\partial_t\tilde{H}+\tilde{\mathbf{V}}\cdot \boldsymbol{\nabla}\tilde{H}$, together with Eq. \eqref{Euler_Eq} and Eq. \eqref{V_and_P}, we arrive at the closed equations
\begin{eqnarray}
    \partial_t \tilde{\mathbf{V}}+\tilde{\mathbf{V}}\cdot \boldsymbol{\nabla}\tilde{\mathbf{V}}+g \boldsymbol{\nabla}\tilde{H}=0\, ,
\end{eqnarray}
and
\begin{eqnarray}
    \partial_t \tilde{H}+\boldsymbol{\nabla} \cdot\big(\tilde{H} \tilde{\mathbf{V}}\big)=0\, .
\end{eqnarray}

To facilitate further analysis, we perform a perturbation expansion around the stationary background $(\mathbf{V}, H)$ by writing
\begin{eqnarray}
    \begin{aligned}
    \tilde{\mathbf{V}}&=\mathbf{V}+\epsilon \mathbf{v}+O(\epsilon^2)\, , \qquad \tilde{H}= H+\epsilon h+O(\epsilon^2)\, ,
    \end{aligned}
\end{eqnarray}
where $H\approx\text{constant}$ under the shallow water approximation. Up to the first-order, the perturbation equations are expressed as
\begin{eqnarray}\label{momentum_Eq}
    \mathrm{D}_t \mathbf{v}+(\mathbf{v} \cdot \boldsymbol{\nabla}) \mathbf{V}+g \boldsymbol{\nabla} h=0\, ,
\end{eqnarray}
and
\begin{eqnarray}\label{H_Eq}
    \mathrm{D}_t h+H \boldsymbol{\nabla} \cdot \mathbf{v}=0\, ,
\end{eqnarray}
where $\mathrm{D}_t=\partial_t+\mathbf{V}\cdot\boldsymbol{\nabla}$ is the material derivative.

We first focus on the standard DBM, which provides an established framework for mimicking Schwarzschild or Kerr black hole geometries. In this model, both the background and perturbation fluid velocities are curl-free, satisfying $\boldsymbol{\nabla}\times\mathbf{V}=0$ and $\boldsymbol{\nabla}\times\mathbf{v}=0$. Thus, the velocity perturbation can be  be expressed in terms of a scalar potential as $\mathbf{v} = \boldsymbol{\nabla} \Psi$. Substituting this into Eq. \eqref{momentum_Eq} and Eq. \eqref{H_Eq} yields the wave equation for $\Psi$
\begin{eqnarray}
    \mathrm{D}_t^2 \Psi-c^2\boldsymbol{\nabla}^2 \Psi=\frac{1}{\sqrt{-\tilde{g}}} \partial_\mu\Big(\sqrt{-\tilde{g}} g^{\mu \nu} \partial_\nu \Psi\Big)=0\, ,
\end{eqnarray}
where $g_{\mu\nu}$ is the analogue spacetime metric, $\tilde{g}$ is its determinant and $c=\sqrt{gH}$ is the speed of sound. The expression after the first equal sign in the above equation implies that the wave equation can be characterized by an effective analogue metric which can be written as
\begin{eqnarray}\label{metric_origin}
    \mathrm{d}s^2&=&g_{\mu \nu} \mathrm{d}x^\mu \mathrm{d} x^\nu \nonumber\\
    &=&-(gH-\mathbf{V}^2) \mathrm{d} \tilde{t}^2-2 \mathbf{V} \cdot \mathrm{d}\mathbf{x} \mathrm{d} \tilde{t}+\mathrm{d}\mathbf{x}^2\\
    &=&-c^2 \mathrm{d} \tilde{t}^2+\Big(\mathrm{d} r-V_r \mathrm{d} \tilde{t}\Big)^2+\Big(r \mathrm{d}\tilde{\theta} -V_\theta \mathrm{d} \tilde{t}\Big)^2\, ,\nonumber
\end{eqnarray}
To eliminate the cross-terms and cast the metric into a more standard form, we introduce the following coordinate transformations
\begin{eqnarray}
    \mathrm{d} t=\mathrm{d} \tilde{t}+\frac{V_r \mathrm{d}r}{c^2-V_r^2}\, , \quad \mathrm{d} \theta=\mathrm{d} \tilde{\theta}+\frac{V_r V_\theta \mathrm{d} r}{r(c^2-V_r^2)}\, .
\end{eqnarray}
Under these new coordinates, the metric \eqref{metric_origin} can simplify to
\begin{eqnarray}\label{metric}
    \mathrm{d}s^2 = -\left(c^2 - V_r^2 - V_\theta^2\right) \mathrm{d}t^2 + \frac{c^2}{c^2 - V_r^2} \mathrm{d}r^2 - 2 r V_\theta \mathrm{d}t \mathrm{d}\theta + r^2 \mathrm{d}\theta^2\, .
\end{eqnarray}
In the standard DBM, the radial and azimuthal components of the fluid velocity are given by
\begin{eqnarray}
    V_r(r)=-\frac{D}{r}\, , \qquad V_\theta(r)=\frac{B}{r}\, .
\end{eqnarray}
By substituting these velocity profiles into the metric~\eqref{metric}, the event horizon can be calculated as
\begin{eqnarray}\label{horizon}
    r_{\text{H}}=\frac{D}{c}\, .
\end{eqnarray}

While the irrotational fluid assumption is a common starting point, it is inadequate for a comprehensive study of exotic resonances. It is therefore natural to incorporate the effects of vorticity. Notably, while vorticity significantly modifies the dynamics of scalar perturbations, its contribution to the background analogue metric \eqref{metric} remains negligible. For the background flow, the vorticity can be characterized by a scalar quantity, defined as twice the intrinsic angular velocity of the fluid elements. Specifically, in our two-dimensional setup, the background vorticity $\Omega_v$ is given by the $z$-component of the velocity curl
\begin{eqnarray}
    \Omega_v=\mathbf{e}_z\cdot\big(\boldsymbol{\nabla}^{(3)}\times\mathbf{V}\big)=-\tilde{\partial}_iV_i=\frac{1}{r}\frac{\partial(rV_\theta(r))}{\partial r}\, ,
\end{eqnarray}
Here, the differential operator $\tilde{\partial}_{i} \equiv \epsilon_{i j} \partial_{j}$ is introduced to represent the two-dimensional curl, where $\epsilon_{ij}$ denotes the Levi-Civita symbol. By construction, this derivative satisfies the condition $\partial_{i} \tilde{\partial}_{i}=\tilde{\partial}_{i} \partial_{i}=0$. Utilizing this operator, we can perform a Helmholtz-like decomposition to express the velocity perturbation, which now incorporates vorticity, in terms of two independent scalar fields $\Psi$ and $\Phi$
\begin{eqnarray}
    \mathbf{v}=\boldsymbol{\nabla}\Psi+\boldsymbol{\tilde{\nabla}}\Phi\, .
\end{eqnarray}
Then Eq.~\eqref{momentum_Eq} becomes as
\begin{eqnarray}
    \partial_i(\mathrm{D}_t \Psi-\Omega_v \Phi+g h)+\tilde{\partial}_i(\mathrm{D}_t \psi+\Omega_v \Psi)+\Phi \partial_i \Omega_v-\Psi \tilde{\partial}_i \Omega_v+(\partial_i V_j) \tilde{\partial}_j \Phi-(\tilde{\partial}_i V_j) \partial_j \Phi=0\, .
\end{eqnarray}
To push the analysis further, we invoke the rigid-body rotation approximation where $\Omega_v \approx \text{constant}$. Consequently, the perturbation equations can be divided into the following form
\begin{eqnarray}
    \begin{aligned}
    \mathrm{D}_t \Psi-\Omega_v \Phi+g h  =0\, ,\qquad
    \mathrm{D}_t \Phi+\Omega_v \Psi  =0\, ,\qquad 
    \mathrm{D}_t h+H \boldsymbol{\nabla}^2 \Psi  =0\, ,
\end{aligned}
\end{eqnarray}
By eliminating the variables $\Phi$ and $h$ from the coupled equations, we arrive at a single, generalized wave equation for the field $\Psi$ that explicitly incorporates the background vorticity
\begin{eqnarray}\label{Tru_WE}
    \mathrm{D}_t^2 \Psi-c^2 \boldsymbol{\nabla}^2 \Psi+\Omega_v^2 \Psi=0\, .
\end{eqnarray}
Given the axial symmetry of the DBM, the wave function $\Psi$ can be naturally decomposed into the angular modes as
\begin{eqnarray}
    \Psi(t,r,\theta)=\sum_m \frac{\mathrm{e}^{\mathrm{i}m\theta}}{\sqrt{r}}\Psi_m(t,r)\, .
\end{eqnarray}
Substituting this ansatz into Eq. \eqref{Tru_WE} and incorporating the DBM vortex profile $V_r$, the wave equation reduces to
\begin{eqnarray}
    \Bigg[ \Big( \frac{\partial}{\partial t} + \frac{\mathrm{i} m V_\theta}{r} \Big)^2 - \frac{\partial^2}{\partial x^2} + V_0(x) \Bigg] \Psi_m(t,x) = 0 \, ,
\end{eqnarray}
where the effective potential $V_0(r)$ is given by
\begin{eqnarray}
    V_0(r) = \Big( c^2 - \frac{D^2}{r^2} \Big) \Big( \frac{m^2 - 1/4}{r^2} + \frac{5D^2}{4c^2r^4} + \frac{\Omega_v^2}{c^2} \Big)\, .
\end{eqnarray}
Finally, transforming the above equation into the frequency domain yields the master equation~\eqref{master_equation}.

In the DBM with vorticity, the fluid naturally divides into two regions. This division is evident in the $\theta$-direction velocity
\begin{eqnarray}
    V_\theta=Br\Theta(r_0-r)+\frac{B}{r}\Theta(r-r_0)\, ,
\end{eqnarray}
where $\Theta(r)$ is the Heaviside step function. This velocity profile implies that $r<r_0$ is the vorticity region. This $\theta$-direction velocity is not ideal for numerical calculations. Therefore, a common form includes a transition region, represented as~\cite{Patrick:2018orp}
\begin{eqnarray}
    V_\theta(r)=\frac{Br}{r_0^2+r^2}\, .
\end{eqnarray}

\section{Coefficient functions of Eq. \eqref{numerical_Eq}}\label{sec:coefficients}
In this appendix, we give the expressions of coefficients in Eq. (\ref{numerical_Eq}) to enhance the readability of this study. The coefficients are given by the following forms
\begin{eqnarray}
    C_2(\sigma) &=& \frac{4 c^4 \sigma (\sigma-2)^2 (\sigma-1)^2}{D^2}\, ,
\end{eqnarray}
\begin{eqnarray}
    C_1^{[1]}(\sigma) &=& -\frac{4 \mathrm{i} c^2 (\sigma-2)^2 (1+\sigma^2)}{D}\, ,
\end{eqnarray}
\begin{eqnarray}
    C_1^{[0]}(\sigma) &=& \frac{4 c^4 (\sigma-2)(\sigma-1)}{D^4 + c^2 D^2 r_0^2} \Big[ \mathrm{i} B D m (\sigma-2)(\sigma-1) + 2 c^2 r_0^2 (1 + 2\sigma(\sigma-2))  + D^2 (2 + 4\sigma(\sigma-2)) \Big]\, ,
\end{eqnarray}
\begin{eqnarray}
    C_0^{[2]}(\sigma) &=& 4 - \sigma(\sigma-1)^2\, ,
\end{eqnarray}
\begin{equation}
\begin{aligned}
    C_0^{[0]}(\sigma) =& \; \frac{c^4}{(D^3 + c^2 D r_0^2)^2 (D^2 + c^2 r_0^2 (\sigma-1)^2)^4} \times \Bigg\{ 2 \mathrm{i} B D^{11} m (\sigma-2)(\sigma-1) (3\sigma-5) + 2 \mathrm{i} B c^{10} D m r_0^{10} (\sigma-2) (\sigma-1)^9 (3\sigma-5) \\
    &\quad + 2 \mathrm{i} B c^8 d^3 m r_0^8 (\sigma-2) (\sigma-1)^7 (3\sigma-5) \big[5 + \sigma(\sigma-2)\big]  + 4 \mathrm{i} B c^6 D^5 m r_0^6 (\sigma-2) (\sigma-1)^5 (3\sigma-5) \big[5 + 2\sigma(\sigma-2)\big] \\
    &\quad + 4 \mathrm{i} B c^4 D^7 m r_0^4 (\sigma-2) (\sigma-1)^3 (3\sigma-5) \big[5 + 3\sigma(\sigma-2)\big]  + 2 \mathrm{i} B c^2 D^9 m r_0^2 (\sigma-2) (\sigma-1) (3\sigma-5) \big[5 + 4\sigma(\sigma-2)\big] \\
    &\quad + D^{12} (\sigma-2) \big[4 + 4m^2 + 5\sigma(\sigma-2)\big] + c^{12} r_0^{12} (\sigma-2) (\sigma-1)^8 \big[4 + 4m^2 + 5\sigma(\sigma-2)\big] \\
    &\quad + D^{10} \Big( -B^2 m^2 (\sigma-4) (\sigma-1)^2 + 2 c^2 r_0^2 (\sigma-2) \big[3 + 2\sigma(\sigma-2)\big] \big[4 + 4m^2 + 5\sigma(\sigma-2)\big] \Big) \\
    &\quad + c^2 D^8 r_0^2 \Big( -4 B^2 m^2 (\sigma-1)^2 \big[-8 + \sigma (11 + \sigma(\sigma-6))\big] \\
    &\qquad\qquad\quad + c^2 r_0^2 (\sigma-2) \big[4 + 4m^2 + 5\sigma(\sigma-2)\big] \big[15 + 2\sigma(\sigma-2) (10 + 3\sigma(\sigma-2))\big] \Big) \\
    &\quad + c^8 D^2 r_0^8 (\sigma-1)^6 \Big( 2 c^2 r_0^2 (\sigma-2) \big[3 + \sigma(\sigma-2)\big] \big[4 + 4m^2 + 5\sigma(\sigma-2)\big] \\
    &\qquad\qquad\qquad\quad + B^2 \big[16(\sigma-2) - m^2 (\sigma(\sigma-4) + 5) (-4 + \sigma(\sigma(\sigma-4) + 5))\big] \Big) \\
    &\quad + 2 c^4 D^6 r_0^4 (\sigma-1)^2 \Big( 2 c^2 r_0^2 (\sigma-2) \big[4 + 4m^2 + 5\sigma(\sigma-2)\big] \big[5 + \sigma(\sigma-2) (5 + \sigma(\sigma-2))\big] \\
    &\qquad\qquad\qquad\quad\;\; + B^2 \big[8(\sigma-2)(\sigma-1)^4 + m^2 (36 + \sigma(-103 + \sigma(124 + \sigma(-76 - 3\sigma(\sigma-8)))))\big] \Big) \\
    &\quad + c^6 D^4 r_0^6 (\sigma-1)^4 \Big( c^2 r_0^2 (\sigma-2) \big[4 + 4m^2 + 5\sigma(\sigma-2)\big] \big[15 + \sigma(\sigma-2)(10 + \sigma(\sigma-2))\big] \\
    &\qquad\qquad\qquad\quad\;\; - 4 B^2 \big[-8(\sigma-2)(\sigma-1)^2 + m^2 (-16 + \sigma(39 + \sigma(-44 + \sigma(26 + \sigma(\sigma-8)))))\big] \Big) \Bigg\}\, .
\end{aligned}
\end{equation}
\bibliography{reference}
\bibliographystyle{apsrev4-1}

\end{document}